\documentstyle[prl,aps,twocolumn,epsfig]{revtex}

\begin{document}

\twocolumn[\hsize\textwidth\columnwidth\hsize\csname @twocolumnfalse\endcsname

\title{Microwave study of quantum $n$-disk scattering}
\author{Wentao Lu, Lorenza Viola, Kristi Pance, Michael Rose, and S. Sridhar}
\address{Department of Physics \\
Northeastern University, Boston, Massachusetts 02115}

\date{\today}

\maketitle

\begin{abstract}
We describe a wave-mechanical implementation of classically chaotic 
$n$-disk scattering based on thin 2-D microwave cavities. Two, three, 
and four-disk scattering are investigated in detail. The experiments,
which are able to probe the stationary Green's function of the system, 
yield both frequencies and widths of the low-lying quantum resonances. 
The observed spectra are found to be in good agreement with calculations
based on semiclassical periodic orbit theory. Wave-vector autocorrelation 
functions are analyzed for various scattering geometries, the small 
wave-vector behavior allowing one to extract the escape rate from 
the quantum repeller. Quantitative agreement is found with the value
predicted from classical scattering theory. For intermediate energies, 
non-universal oscillations are detected in the autocorrelation
function, reflecting 
the presence of periodic orbits.
\end{abstract}


\vskip1.5pc]

\section{Introduction}

The system of $n$-disks on a plane is perhaps the simplest and 
paradigmatic example of a chaotic scattering system 
\cite{Eckhardt87,Cvitanovic88,Gaspard89,Gaspard92,Gaspard98}. 
The classical differential cross-section of the system with $n>2$ disks
on a plane is singular, with singular points forming a Cantor set
\cite{Eckhardt87,Gaspard89}. Thus, the investigation of chaotic scattering 
geometries naturally brings in the capability to access the intrinsic 
{\sl fractal} nature of the underlying classical repeller.

From the perspective of quantum-classical correspondence, the 
wave-mechanical counterpart of this ``fractal pinball game'' is even 
more intriguing. Semiclassical attempts to treat this problem led to 
important theoretical advances and to the development of sophisticated 
semiclassical tools, notably cycle expansion, that represents one of the
more productive applications of periodic orbit theory 
\cite{Gaspard98}.

From a broader perspective, the $n$-disk scattering can be regarded as 
the prototype of an {\sl open quantum chaos system. }
For closed quantum chaotic systems (billiards in particular), theoretical
and experimental results are available on spectral eigenvalues (which are
purely real), and a good understanding of both universal and non-universal
features has been reached \cite{Bohigas90}. Open chaotic systems are a
special class of open quantum systems, whereby bound states of a closed
geometry are converted to long-lived metastable states due to coupling to
continua. Accordingly, eigenvalues are intrinsically complex and their
universal behavior is currently a subject of considerable interest
\cite{John91}.

Investigation of $n$-disk scattering was originally stimulated by the attempt 
of modelling unimolecular reactions \cite{Gaspard89}. However, the system 
turns out to be a good exemplification for a variety of physical situations,
ranging from crossroad geometries for semiconductor devices,
to electromagnetic and acoustic scattering, and heavy-ion nuclear reactions 
\cite{chaos93}.

In spite of extensive theoretical analysis, there have been few direct 
experimental implementations of such chaotic model geometries so far. 
In this paper we present the results of a systematic experimental study of
the quantum resonances and decay characteristic of 2-D scattering
repellers. The experiments utilize thin microwave geometries where, under 
appropriate conditions, the underlying Helmholtz equation maps exactly into 
the time-independent Schr\"{o}dinger equation in two dimensions. 
Experiments were carried out for $n=1,2,3,4$, as well as for large $n=20$, 
the latter corresponding to the so-called random Lorentz scatterer. 
A concise account of the main results has been presented in \cite{Lu99}. 

The experiments, which are able to access the stationary Green's function of
the system, yield the frequencies and the widths of the low-lying quantum 
resonances. We carried out semiclassical calculations of such resonances, 
which are found to reproduce the measured spectra reasonably well. Our experiments 
enable us to explore the role of symmetry in a unique way by probing different 
irreducible representations of the symmetry group of the scatterer. The experimental
data are used to identify both universal and non-universal signatures of the 
classical chaos in the transmission spectra, through measures such as the spectral
(wave-vector) autocorrelation function.

Correlation functions are a valuable tool to extract key information on the
spectral properties of the system. In mesoscopic conductors, measurements of
the magnetic field correlation of the conductivity have provided unique
insight into the manifestations of the chaotic classical dynamics at the
quantum level \cite{Jalabert90}. Correlations of wave functions in chaotic
systems have been considered recently as a probe for classical ergodicity
predictions \cite{Dorr99}. In the present experiments, we take advantage of
our ability to vary the wave-vector of the system to directly access energy
correlations, that are difficult to extract in semiconductor
microstructures. The small $k$ (long time) behavior of the resulting
autocorrelation provides a measure of the {\sl quantum escape rate}, which is
shown to be in good agreement with the corresponding classical value.
For large $k$ (short time), the contribution of periodic orbits is evidenced
through non-universal oscillations of the autocorrelation function.

The content of the paper is organized as follows. After describing the
experimental realization in Sec. II, the connections between
electromagnetic and quantum mechanical scattering, and between the observed
transmission amplitude and the stationary Green's function are elucidated in
Sec. III. In Sec. IV, the essential background on semiclassical theory is
recalled, with emphasis on the role played by symmetry properties. The
remaining sections are devoted to the presentation and the interpretation of
the experimental results: in Sec. V, we present a comparison between the
measured traces and the semiclassical predictions for various scattering
geometries probed in the experiment, while in Sec. VI we focus on the
analysis of spectral autocorrelations, from which the wave-mechanical escape
rate can be extracted. An overall discussion of the results and their
implications, together with an outlook on open issues and future work are
presented in the conclusive section.

\section{Experimental realization}

The experiments were carried out in thin microwave structures consisting of
two highly conducting Cu plates spaced $d\,\simeq 6\,$mm and about 
$55\times 55$ cm$^{2}$ in area. 
A picture of a typical experimental setup is displayed in Fig. 1.
Similar experiments in {\sl closed} geometries have provided direct observation
of scars \cite{Sridhar91}, enabled precise tests of 
eigenvalue statistics 
\cite{Kudrolli94}, and allowed experimental studies of localization effects
\cite{Kudrolli95a}. Disks and bars also made of 
\hspace{0pt}Cu and of thickness $d$ were placed between the plates and in
contact with them. Disks of radius $a=2$ and $a=5$ cm were used in the
experiments. In order to simulate an infinite system, microwave absorber
material was sandwiched between the plates at the edges. Microwaves were 
fed into the system by inserting a loop-terminating coaxial line in the
vicinity of the scatterers. Since we ensure that the relation 
$f<f_{c}=c/2d=25$ GHz is valid for all operating frequencies (here 
$k=2\pi f/c $, $c$ being the speed of light), 
the only allowed electromagnetic modes are
a class of TM (Transverse Magnetic) modes with no variation along the
direction $z$ orthogonal to the plates. The nonvanishing field components
for these states are the axial component of the electric field $E_{z}$, and
transverse components of the magnetic field $H_{x},H_{y},$ or 
${\bf H}\ =(1/ik)\,\hat{z}\times \nabla E_{z}$. 
Thus, essentially, the geometry of the
experiment can be regarded as 2-dimensional, the whole field configuration
being accessible from knowledge of $E_{z}$ alone. The output signal can be
picked up by both directly coupling to the axial electric field $E_{z}$ 
({\sl electric coupling}) or by measuring the associated transverse magnetic
field ${\bf H}$ ({\sl magnetic coupling}). In the first configuration, a
microwave pin is inserted through appropriate holes drilled on the top plate
of the cavity, while magnetic coupling is established by suitably
positioning a second loop perpendicular to the plates. Note
that, for a sufficiently small loop area, the measured signal is simply
proportional to the electric field value at the pick-up probe location in
both cases. Data from both electric and magnetic coupling configurations are
used in the subsequent analysis.

The resonances were probed in the transmission mode by sweeping the frequency
in the range between 0 and 20 GHz. All the measurements are carried out
using an HP8510B vector network analyzer measuring the complex transmission
coefficient $S_{21}(f)\equiv (X+iY)(f)$ of the coax plus scatterer
system. It is crucial
to ensure that there is no spurious background scattering due to the finite
size of the system. This was verified carefully as well as that the effects
of the coupling probes were minimal and did not affect the results. Also, we
stress that dissipation effects due to finite conductivity of the walls are
entirely negligible in the present experiments. 

\section{Green's function and experimental $S$-parameters}

\subsection{Electromagnetic vs. quantum mechanical scattering}

According to Maxwell equations, the problem of stationary scattering of
electromagnetic waves from perfect metallic conductors in a 2-D plane is
defined mathematically by the Helmholtz equation for the field component

\begin{equation}
\left( \nabla ^{2}+k^{2}\right) E_{z}(\vec{r})=0,  \label{helmoltz}
\end{equation}
for $\vec{r}=(x,y)$ outside the scattering region ({\it i.e.}, between the
disks), supplemented by Dirichlet boundary conditions on the perimeter of
any scatterer:

\begin{equation}
E_{z}(\vec{r})=0,\quad \vec{r}\in {\rm perimeters},
\end{equation}
implying, for any geometry, a vanishing electric field inside the disks. The
Helmholtz equation is formally identical to the time-independent
Schr\"{o}dinger equation with $\Psi =E_{z}$. Since, in addition, the same
boundary conditions apply to both the axial electric field and the
wavefunction, the problem is equivalent to quantum mechanical scattering
from hard disks. This correspondence is exact in the whole range of
frequencies below the cutoff frequency $f_{c}$ that are considered
in the experiment.

According to quantum scattering theory, the stationary properties of the
scattering process are characterized completely by the collection of 
$S$-matrix elements, expressing the transition amplitudes between asymptotic
incoming and outgoing states \cite{Gaspard98,newton}. The $S$-matrix
exhibits poles as a function of the wave vector (or energy) of the incoming
wave. These poles, corresponding to {\sl scattering resonances}, occur in
the complex $k$-plane, $k=k^{\prime }+ik^{\prime \prime }$. The imaginary
part is interpreted as a resonance width, implying that scattering
resonances are associated with {\sl metastable }states having a finite
lifetime. Note that, in contract to closed-cavity geometries where non-zero
widths only arise due to the finite wall conductivity \cite{balam}, widths
have here a purely {\sl geometric} origin, the openness of the system
preventing a stable trapping of the states within the scattering geometry
for arbitrarily long times. We shall use the convention $k=s-i\gamma$ for a
given resonance, {\it i.e.}, $s=k^{\prime }$ to denote the position of the
resonance along the $k$-axis and 
$\gamma =-\,k^{\prime \prime },\,k^{\prime \prime }<0$ 
to denote half-widths in the wave-vector domain. $\gamma$ is
related to the observed resonance width $\Delta f$ in the frequency spectrum
by $\Delta f=(c/2\pi )\,\,\Delta k=(c/2\pi )\,\,2\gamma$.

The correspondence, pointed out above, between the equations of 
motion and the
boundary conditions satisfied by electromagnetic ({\sc em}) and
quantum-mechanical ({\sc qm}) stationary scattering allows one to establish
a direct mapping between the $S$-matrix spectral properties in the domain of 
{\sl complex wave vectors}:
\begin{equation}
\left( k^{\prime }+ik^{\prime \prime }\right) _{\text{{\sc em}}}
=(k^{\prime}+ik^{\prime \prime })_{\text{{\sc qm}}}.  \label{complex}
\end{equation}
This mapping enables us to study the quantum properties of the $n$-disk 
system. Note that quantum-mechanical (half) widths $\Gamma/2 $ for unstable 
states are typically defined in the {\sl energy} domain as 
$E=E^{\prime }+iE^{\prime \prime }=\varepsilon -i\Gamma/2,\, 
E^{\prime \prime}<0$, the relation between (complex) pseudo-energies and 
(complex) wave-vectors being $E=(\hbar k)^{2}/2m$ as in the real case 
\cite{datta}, \cite{fyodorov}. Thus, 
$E^{\prime \prime }=(\hbar ^{2}/m)\,k^{\prime}k^{\prime \prime}$. 
However, due to the difference between the non-stationary propagators for 
{\sc em} and {\sc qm} problems, care should be taken when introducing and 
comparing associated quantities like decay rates or lifetimes which are 
defined in the time domain. By exploiting the time-dependent Schr\"{o}dinger 
equation, quantum decay rates for square amplitudes $|\Psi (t)|^{2}$ are 
found as inverse lifetimes $1/\tau =\Gamma/\hbar =|2E^{\prime \prime}|/\hbar$. 
One gets $\tau \cdot \Gamma /\hbar =1$, expressing Heisenberg's uncertainty
principle. From the full wave equation for a monochromatic field, the decay
rate of the field amplitude $|E_{z}(t)|^{2}$ is given by 
$1/\tau =2\gamma c=c|2k^{\prime \prime }|$. Accordingly, $\tau \cdot 2\gamma c=1$. 
In spite of such a difference, the wave mechanical spectrum will allow
us to extract a quantity (the classical escape rate) which is related to the 
{\sl average} trapping time of the underlying classical trajectories inside
the scattering region.

\subsection{Green's function}

We turn now to outline the relationship between the measured response  
$S_{21}$ and the two-point Green function of \thinspace the chaotic billiard.
The equation of motion satisfied by the relevant field component $E_{z}$ in
the presence of a monochromatic driving at frequency $\omega$ is
\begin{equation}
\left( \nabla ^{2}-\frac{1}{c^{2}}\frac{\partial ^{2}}{\partial t^{2}}
\right) E_{z}(\vec{r},t)=E_{0}(\vec{r})\,e^{-i\omega t},
\end{equation}
$\vec{r}$ denoting 2-D coordinates in the plane and $E_{0}(\vec{r})$ being 
the associated amplitude for waves fed into the cavity with wave vector 
$k=\omega /c$. For a stationary field configuration, the above equation reads

\begin{equation}
(\nabla ^{2}+k^{2})E_{z}(\vec{r})=E_{0}(\vec{r}).
\end{equation}

By definition, the so-called Green's function (or propagator) associated
with the wave equation differential operator satisfies the equation

\begin{equation}
\left( \nabla ^{2}-\frac{1}{c^{2}}\frac{\partial ^{2}}{\partial t^{2}}
\right) G(\vec{r},\vec{r}_{0};t,t_{0})=\delta (\vec{r}-\vec{r}_{0})\delta
(t-t_{0})\,,
\label{green}
\end{equation}
for fixed $\vec{r}_{0}$ and $t_{0}.$ The Green's function 
$G(\vec{r},\vec{r}_{0};t,t_{0})$ is identical to the wavefunction generated 
by the system at point $\vec{r}$ and time $t$ in response to a delta-like 
excitation applied at point $\vec{r}_{0}$ and time $t_{0}$. 
For a time-homogeneous system, 
$G(\vec{r},\vec{r}_{0};t,t_{0})=G(\vec{r},\vec{r}_{0};t-t_{0})$, 
and Fourier transformation of Eq. (\ref{green}) gives

\begin{equation}
\left( \nabla ^{2}+k^{2}\right) \,G(\vec{r},\vec{r}_{0};k)
=\delta (\vec{r}-\vec{r}_{0})\,,
\end{equation}
where the stationary Green's function has been introduced:

\begin{equation}
G(\vec{r},\vec{r}_{0};t-t_{0})=\frac{1}{2\pi }
\int G(\vec{r},\vec{r}_{0};k)\,e^{ick(t-t_{0})}\,dk\,.
\end{equation}
Knowledge of the Green's function allows one to calculate the response of
the system to a generic excitation $E_{0}(\vec{r})$ as a convolution
integral 
$E_{z}(\vec{r}_{2})=\int_{V}G(\vec{r}_{2},\vec{r};k)E_{0}(\vec{r})\,d\vec{r}$. 
In particular, if a point-like probe is placed at $\vec{r}_{1}$
to excite the system, 
$E_{0}(\vec{r})=E_{0}(\vec{r}_{1})\,\delta (\vec{r}- \vec{r}_{1})$, 
we simply get 
$E_{z}(\vec{r}_{2})=G(\vec{r}_{2},\vec{r}_{1};k)\,E_{0}(\vec{r}_{1})$.

The connection with the observable quantity of our experiments is
established by recalling that the transmission coefficient 
$S_{21}=V_{out}/V_{in},$ where $V_{in}$ and $V_{out}$ are
linear responses to the electric field at the probe locations $\vec{r}_{1}$
(input) and $\vec{r}_{2}$ (output) respectively. Note that $S_{21}$ can be
related to the $S$-matrix element describing transmission between waves
entering through antenna 1 and going out through antenna 2 \cite{Stein95}.
We have $V_{out}=\alpha E_{z}(\vec{r}_{2})$, 
$V_{in}=\beta E_0 (\vec{r}_{1})$, $\alpha$ and $\beta$ 
denoting impedance factors that are characteristic of the coax lines 
and the analyzer, and are generally slowly-varying functions of 
frequency. Thus we get 
$S_{21}=(\alpha /\beta)\,G(\vec{r}_{2},\vec{r}_{1};k)$, 
which we can write in the form
\begin{equation}
S_{21}(k)=A(k)\,\,G(\vec{r}_{2},\vec{r}_{1};k)\,.
\end{equation}
In a formal analogy with the closed-system case, the Green's function can be
expressed in terms of a generalized eigenfunction expansion \cite{datta},
\begin{equation}
G(\vec{r}_{2},\vec{r}_{1};k)=\sum_{n}\frac{\Psi_{n}(\vec{r}_{2})
\Phi_{n}^{*}(\vec{r}_{1})}{k_{n}^{2}-k^{2}}\,,  \label{expansion}
\end{equation}
$\Psi_{n}(\vec{r})$ denoting the $n$-th eigenfunction of an effective
open-system Hamiltonian, with corresponding adjoint eigenstate 
$\Phi_{n}(\vec{r})$ and associated (complex) wave vector $k_{n}$. It is worth
stressing that the measured Green's function is identical with the actual
Green's function of the open system provided the perturbation introduced by
the probes is negligible. If this is not the case, a modified Green's
function is generally obtained, with {\sl shifted} complex poles in the
expansion (\ref{expansion}) ({\it i.e.}, resonances are shifted and
broadened due to the presence of the probes). In the experiment, we ensured
that the antennas were only {\sl weakly} coupled to the system, with
associated perturbation effects less than $\approx 10^{-3}$ of the observed
frequencies and widths. Under these conditions, the relation established
above between $S_{21}$ and the Green's function generalizes the
derivation for closed microwave billiards presented in \cite{Stein95}.

\subsection{Comparison with mesoscopic conductivity measurements}

The connection between so-called transmission function, $S$-matrix, and
Green's function is well known in the field of mesoscopic transport 
\cite{datta}. In particular, we recall that the {\sl transmission function} 
of a coherent conductor between two leads 1, 2 is defined as \cite{Marcus92} 
\begin{equation}
T=\sum_{m\in 1,n\in 2}|t_{mn}|^{2},
\end{equation}
where the transmission coefficient $t_{mn}$, which characterizes the
transmission amplitude between mode $m$ in lead 1 and mode $n$ in lead 2, is
given by
\begin{equation}
t_{mn}=-i\hbar \sqrt{v_{m}v_{n}}\int dy'dy\phi_{m}^*(y')\phi_{n}(y)
G(y^{\prime },x_2,y,x_{1};k).
\label{delta}
\end{equation}
Here, $v_{m}(v_{n})$ and $\phi_{m}(\phi_{n})$ are the longitudinal
velocity and the transverse wave function for the mode $m$ in lead 1 
($n$ in lead 2) respectively, while $x_{1(2)}$ denote the longitudinal
coordinate of antennas 1, 2. The transmission coefficients, which are
clearly energy (frequency) dependent, are directly related to the
correspondent elements of the $S$-matrix. In fact, the above expression for 
$t_{mn}$ is a two-probe version of the general connection known as the
Fisher-Lee relation between the $S$-matrix and the Green's function 
\cite{fisher}. The transmission $T$ is 
related directly to the {\sl conductance} $\sigma $ of the 
conductor by the Landauer formula, $\sigma =(e^{2}/h)\,T$, 
$e$ denoting the electron charge \cite{datta}.

In the case of a point-like excitation and a point-like probing of the
system, the input/output leads act as zero-dimensional tunneling point
contacts. Thus, the transverse dimension of the leads can be neglected and
the wave functions $\phi _{m}(y)$ in Eq. (\ref{delta}) are proportional to 
$\delta$-functions, 
$\phi_{m}(y^{\prime })=\delta (y^{\prime }-y_{2}),\phi_{n}(y)=\delta (y-y_{1})$. 
By combining the two expressions above, and
putting $\vec{r}_{1}=(x_{1},y_{1})$, $\vec{r}_{2}=(x_{2},y_{2})$, we get
$T(k)\propto |G(\vec{r}_{2},\vec{r}_{1},k)|^{2}$. 
Since the Green's
function is directly related, as shown in the previous section, to the
experimental trace, one can interpret

\begin{equation}
T(k)\propto |S_{21}(k)|^{2}
\end{equation}
as a measurement of the two-point conductance. As a consequence, direct
comparison is possible with theories originally developed for electronic
micro-structures \cite{Jalabert90,Marcus92,Jensen96,Delos96}. Note that in
the language of mesoscopic conductors, the weak-coupling assumption between
the system and the point-like leads implies the possibility of neglecting,
as noticed above, any perturbation effect associated with the so-called lead
self-energy \cite{datta}. From a general point of view, microwave-based
implementations offer, compared to their solid-state counterparts, the
advantage of a simple and practically unlimited manipulation of geometrical
properties, along with the possibility of changing the wave-vector (and
thereby the energy) of the incoming waves at will. Lastly, since
electromagnetic waves are intrinsically non-interacting, a direct analogy
with the ideal limit of a non-interacting electron gas applies and microwave
experiments automatically probe quantum transport in the ballistic regime.

\section{Semiclassical theory}

The starting point for the semiclassical derivation is the
concept of generalized density of states $D(E)$ which, according to Balian
and Bloch \cite{Balian74}, is defined as a suitable difference between the
density of states of the free and of the scattering system. The relative
density of states is related to the $S$-matrix in the following way: 
\begin{equation}
D(E)={1\over 2\pi i}{\rm tr}\Bigg[S^{\dagger} {dS(E)\over dE}\Bigg]\,.
\label{density}
\end{equation}
Therefore, the density $D(E)$ and the $S$-matrix share the same complex
poles. $D(E)$ itself is written as
\begin{equation}
D(E)=-\frac{1}{\pi }\,{\rm Im}\,g(E)\,,  \label{density2}
\end{equation}
where the function $g(E),$ which is related to the so-called quantum Selberg
zeta function, is the trace of the difference between the Green's function
in the presence and in the absence of the disks respectively.

In the semiclassical limit, $g(E)$ is expressed as a sum over periodic
orbits {\it i.e., }by Gutzwiller's trace formula \cite{Gutzwiller90}. The
trace function can then be written in terms of the {\sl Ruelle }
$\zeta $-{\sl function} as \cite{Gaspard89},
\begin{equation}
g(E)=g_{0}-\sum_{j=0}^{\infty }{\frac{\partial }{\partial E}}\ln \zeta
_{(1/2)+j}(-ik)\,,  \label{ruelle}
\end{equation}
where $g_{0}$, which is independent on the energy $E$, is given by the
difference between the so-called Thomas-Fermi state densities with and
without the disks, {\it e.g., }$g_{0}=-4a^{2}m/(2\hbar ^{2})$ in the 4-disk
configuration. In Eq. (\ref{ruelle}), the Ruelle $\zeta $-function with $j$
running from $0$ to $\infty $ is given explicitly as
\begin{equation}
\zeta_{(1/2)+j}(-ik)=
\prod_{p}\,\left[1-(-1)^{L_{p}}e^{ikl_{p}}/\Lambda_{p}^{(1/2)+j}\right]^{-1},  
\label{ruelle2}
\end{equation}
where $k$ is the wave vector, $l_{p}$ is the length of the periodic orbit $p$
, $L_{p}$ the number of collisions of the periodic orbit with the disks, and 
$\Lambda_{p}$ the eigenvalue of the so-called {\sl stability} (or {\sl
monodromy matrix}).
According to the above expressions, the task of finding the scattering
resonances of the system is reduced to the problem of identifying the
complex poles of the Ruelle $\zeta$-function. Since the Ruelle $\zeta$
-function with $j=0$ provides the leading contributions to the resonances
with longer lifetimes (sharper resonances), we will only restrict
ourselves to this case in our subsequent discussion.

The Ruelle $\zeta $-function can be conveniently expressed in terms of a
so-called {\sl Euler product representation},
\begin{equation}
\zeta ^{-1}=\prod_{p}\,(1-t_{p})\,,  \label{euler}
\end{equation}
$t_{p}$ denoting the semiclassical weight for cycle $p$ \cite{Cvitanovic93}.
In practice, the periodic orbit formula for the Ruelle function is evaluated
by performing a {\sl cycle expansion} and investigating the zeros and radii
of convergence as functions of truncations to cycles of a given maximum
topological length. In the $n$-disk problem, if the disks are positioned 
sufficiently far from each other along a ring, it is possible to travel
between any three successive of them and the trapped trajectories can be put
in one-to-one correspondence with the bi-infinite sequence of symbols 
$\omega_{k}$ taken in the alphabet $\{1,2,3,\ldots ,n\}$ with  
$\omega _{k+1}\neq \omega _{k}$. The latter constraint implies that the 
topological entropy per bounce is equal to $\ln (n-1)$. Thus, the system 
symbolic dynamics has a finite grammar and the Euler product (\ref{euler}) 
can be rewritten by separating out a dominant fundamental contribution and 
the remaining corrective terms:
\begin{equation}
\prod_{p}\,(1-t_{p})=1-\sum_{f}t_{f}-\sum_{r}c_{r}\,.  \label{euler2}
\end{equation}
Here, the number of fundamental terms $t_{f}$ is equal to the number of
symbols in the unconstrained symbolic dynamics, while the so-called 
{\sl curvature corrections} $c_{r}$ represent contributions due to the
nonuniformity of the system \cite{Cvitanovic93}.

\smallskip 
In our experiments, $n$ identical disks are placed on a ring with
equal space between the nearest neighbors. Accordingly, the system is
characterized by symmetry point group ${\sl G}=C_{2v}$, $C_{3v},$ $C_{4v},$ 
$C_{5v},$ $C_{6v},\ldots $ for $n=2,3,4,$ {\it etc.}, and symmetry properties
can be used to classify the corresponding scattering resonances. In
practice, the presence of symmetries can also be exploited to simplify
the cycle expansion and improve its convergence. The key concept is to
remove symmetry-induced degeneracies between cycles by reducing the dynamics
to the so-called {\sl fundamental domain (FD)}. The latter is a region obtained
by ideally replacing the symmetry axes with perfectly reflecting mirrors.
Global periodic orbits of the full system can be described completely by
folding irreducible segments into periodic orbits in the fundamental domain.
Correspondingly, the Ruelle $\zeta $-function can be factorized in the
product over different irreducible representations ({\sl irreps}) of the
symmetry group \cite{Cvitanovic93}. By taking into account the 
degeneracies of the periodic
orbits and working in the fundamental domain, the Euler product (\ref{euler})
 takes the form

\begin{equation}
\prod_{p}\,(1-t_{p})=\prod_{\tilde{p}}\,(1-t_{\tilde{p}}^{h_{p}})^{m_{p}}\,,
\label{euler3}
\end{equation}
where a modified semiclassical weight has been introduced:

\begin{equation}
t_{\tilde{p}}=t_p^{1/h_p}=
\frac{(-1)^{L_{\tilde{p}}}}{\sqrt{\Lambda _{\tilde{p}}}}\,
\exp(ikl_{_{\tilde{p}}})\,.
\label{tp}
\end{equation}
In Eqs. (\ref{euler3})-(\ref{tp}), $m_{p}=g/h_{p}$, $g$ and $h_{p}$ denoting 
respectively the order of the full symmetry group $G$ and of the maximal 
subgroup leaving $p$ invariant. $\tilde{p}$ is the irreducible segment of 
the orbit $p$ that corresponds to a fundamental domain orbit, 
$L_{\tilde{p}}$ is the number of collisions with the disk in the fundamental
domain and $\Lambda _{\tilde{p}}=$ $\Lambda _{p}^{1/h_{p}}$ the eigenvalue
of the stability matrix in the fundamental domain. The calculation of 
$\Lambda _{\tilde{p}}$ can be easily performed by diagonalizing the stability
matrix $J_{\tilde{p}}$ which, for hard-disk billiards, reads as 
\cite{Cvitanovic93,Gaspard94}

\begin{equation}
J_{\tilde{p}}=(-1)^{n_{_{\tilde{p}}}}\prod_{k=1}^{n_{\tilde{p}}}\,\left( 
\begin{array}{ll}
1 & l_{k} \\ 
0 & 1
\end{array}
\right) \left( 
\begin{array}{ll}
1 & 0 \\ 
r_{k} & 1
\end{array}
\right) \,.
\end{equation}
Here, for simplicity we rescaled the constant velocity to unit value. $l_{k}$
denotes the length of the $k$-th free-flight segment of the cycle 
$\tilde{p} , $ while $r_{k}=2/a\cos \phi _{k}$ is the defocusing due to the $k$-th
reflection, occurring at an incidence angle $\phi _{k}$.

By construction, only one irreducible representation appears whenever the
system is probed in the fundamental domain. Note that, in practice, one can
easily switch from one symmetry to another by varying the angle 
$\theta =\pi/n$. We recall now some relevant formulas for the symmetry 
configurations examined in the experiment.

\subsection{$C_{2}$ factorization}

The group $G=C_{2v}$ is the appropriate symmetry group for two-disk
scattering which, without being chaotic, offers the most important example
of a nontrivial {\sl integrable} scattering problem. An exact analytical
solution is available for the classical dynamics \cite{Jose92}, while a full
quantum mechanical calculation of the scattering resonances has been
performed in \cite{Wirzba92,Decanini98}. From the standpoint of the
semiclassical analysis, the symmetry group of the periodic orbits is $G=C_{2}$,
characterizing the transformation properties of the orbits under the 
exchange of the two disks \cite{Wirzba92}. 
The fundamental domain is the half-space containing a single disk. 
There are just two group elements, $C_{2}=\{e,P\}$, 
$e$ being the identity and $P$ the parity operation, and only one periodic
orbit. Cycles classify according to the two irreducible representations
$A_{1}$(symmetric) and $A_{2}$(antisymmetric). Let $a$ and $R$ denote
the radius and the center-center disk separation respectively, with the ratio
$\sigma \equiv R/a$. We have
\begin{equation}
\zeta_{A_{1}}^{-1}=1-t_{0}\hskip 1cm \zeta_{A_{2}}^{-1}=1+t_{0}\,,
\end{equation}
with $t_{0}=-\exp[ik(R-2a)]/\sqrt{\Lambda}$, and 
\begin{equation}
\Lambda=(\sigma-1)+\sqrt{\sigma(\sigma-2)}\,,  \label{Lambda}
\end{equation}
denoting the eigenvalue of the monodromy matrix indicated above. 

We thus get the semiclassical scattering resonances as
\begin{equation}
k_{n}=\frac{(2n-m)\,\pi -\frac{i}{2}\ln \Lambda }{R-2a}\,,
\qquad n=1,2,\ldots\,,
\label{two}
\end{equation}
where $m=1$ for the $A_{1}$-irrep, $m=0$ for the $A_{2}$-irrep. 
In the fundamental domain, only the antisymmetric $A_{2}$ representation
contributes.

\subsection{$C_{3v}$ factorization}

The symmetric three-disk pinball is invariant under the transformations of
the group $C_{3v}.$ In addition to the identity transformation, they include
two rotations through $2\pi /3$ and $4\pi /3$ about the main axis, and three
mirror reflections around the symmetry axes. The fundamental domain is
bounded by a disk and the two adjacent sections of the symmetry axes acting
as mirrors (1/6th of the full space, see Fig. 3, inset). 
$C_{3v}$ has two 1-D irreps $A_{1}$ and $A_{2}$ (symmetric and antisymmetric 
under reflections respectively), 
and one 2-D irrep of mixed symmetry labelled $E$. The 3-disk
dynamical zeta function factorizes into: 
$\zeta =\zeta _{A_{1}}\zeta_{A_{2}}\zeta _{E}^{2}$,  
the contributions of each given irreducible representation being given by 
the following curvature expansion \cite{Cvitanovic93}:
\begin{eqnarray}
\zeta_{A_{1}}^{-1} &=&1-t_{0}-t_{1}-(t_{01}-t_{0}t_{1}) \nonumber\\
&&-[(t_{001}-t_{0}t_{01})+(t_{011}-t_{1}t_{01})]-\cdots ,  
\end{eqnarray}
for the $A_{1}$subspace,

\begin{eqnarray}
\zeta _{A_{2}}^{-1} &=&1+t_{0}-t_{1}+(t_{01}-t_{0}t_{1}) \nonumber\\
&&-[(t_{001}-t_{0}t_{01})-(t_{011}-t_{1}t_{01})]-\cdots \,,  
\end{eqnarray}
for the antisymmetric $A_{2}$ subspace, and, for the mixed-symmetry subspace 
$E$,

\begin{eqnarray}
\zeta _{E}^{-1} &=&1+t_{1}-(t_{0}^{2}-t_{1}^{2})+(t_{001}-t_{1}t_{0}^{2}) \nonumber\\
&&+[t_{0011}+(t_{001}-t_{1}t_{0}^{2})t_{1}-t_{01}^{2}]+\cdots \,.  
\end{eqnarray}
The representation in the fundamental domain is $A_{2}$. A detailed
comparison between the semiclassical predictions and the exact quantum
resonances for the 3-disk scattering problem has been reported in  
\cite{Gaspard89,Eckhardt95}. Exact quantum calculation was also performed in
\cite{Decanini98}.

\subsection{$C_{4v}$ factorization}

The scattering problem of four equal disks placed on the vertices of a
square is characterized by $C_{4v}$ symmetry. This is a group consisting of
the identity, two reflections across the coordinate axes, two diagonal
reflections, and three rotations by angles $\pi /2,\pi $ and $3\pi /2$. The
fundamental domain is a sector delimited by a disk, a portion of the
corresponding diagonal axis and a portion of the concurrent coordinate axis (
{\it i.e.,} 1/8th of the full space, see Fig. 4b, inset). 
$C_{4v}$ has four 1-D irreps,
either symmetric ($A_{1})$ or antisymmetric ($A_{2})$ under both types of
reflections, or symmetric under one and antisymmetric under the other 
($B_{1} $, $B_{2}$), and one 2-D representation $E$. The $\zeta $-function is
factorized as 
$\zeta =\zeta _{A_{1}}\zeta _{A_{2}}\zeta _{B_{1}}
\zeta_{B_{2}}\zeta _{E}^{2}$, 
where the contributions for the various invariant
subspaces have the following curvature expansion \cite{Cvitanovic93}:

\begin{eqnarray}
\zeta _{A_{1}}^{-1} &=&1-t_{0}-t_{1}-t_{2}  \nonumber \\
&&-(t_{01}-t_{0}t_{1}+t_{02}-t_{0}t_{2}+t_{12}-t_{1}t_{2})  \nonumber \\
&&-(t_{001}-t_{0}t_{01})-(t_{002}-t_{0}t_{02})-(t_{011}-t_{1}t_{01}) 
\nonumber \\
&&-(t_{022}-t_{2}t_{02})-(t_{112}-t_{1}t_{12})-(t_{122}-t_{2}t_{12}) 
\nonumber \\
&&-(t_{012}+t_{021}+t_{0}t_{1}t_{2}-t_{0}t_{12}-t_{1}t_{02}-t_{2}t_{01})
\ldots \,,  \nonumber \\
\zeta _{A_{2}}^{-1} &=&1+t_{0}-t_{1}+(t_{01}-t_{0}t_{1})+t_{02}-t_{12} 
\nonumber \\
&&-(t_{001}-t_{0}t_{01})-(t_{002}-t_{0}t_{02})+(t_{011}-t_{1}t_{01}) 
\nonumber \\
&&+t_{022}-t_{122}-(t_{112}-t_{1}t_{12})  \nonumber \\
&&+(t_{012}+t_{021}-t_{0}t_{12}-t_{1}t_{02})\ldots \,,  \nonumber \\
\zeta _{B_{1}}^{-1} &=&1-t_{0}+t_{1}+(t_{01}-t_{0}t_{1})-t_{02}+t_{12} 
\nonumber \\
&&+(t_{001}-t_{0}t_{01})-(t_{002}-t_{0}t_{02})-(t_{011}-t_{1}t_{01}) 
\nonumber \\
&&-t_{022}+t_{122}-(t_{112}-t_{1}t_{12}) \\
&&+(t_{012}+t_{021}-t_{0}t_{12}-t_{1}t_{02})\ldots \,,  \nonumber \\
\zeta _{B_{2}}^{-1} &=&1+t_{0}+t_{1}-t_{2}  \nonumber \\
&&-(t_{01}-t_{0}t_{1})+(t_{02}-t_{0}t_{2})
+(t_{12}-t_{1}t_{2})  \nonumber\\
&&+(t_{001}-t_{0}t_{01})-(t_{002}-t_{0}t_{02})+(t_{011}-t_{1}t_{01}) 
\nonumber \\
&&+(t_{022}-t_{2}t_{02})-(t_{112}-t_{1}t_{12})+(t_{122}-t_{2}t_{12}) 
\nonumber \\
&&-(t_{012}+t_{021}+t_{0}t_{1}t_{2}-t_{0}t_{12}-t_{1}t_{02}-t_{2}t_{01})
\cdots \,,  \nonumber \\
\zeta _{E}^{-1}
&=&1+t_{2}-(t_{0}^{2}-t_{1}^{2})+(2t_{002}-t_{2}t_{0}^{2}-2t_{112}-t_{2}t_{1}^{2})
\nonumber \\
&&+(2t_{0011}-2t_{0022}+2t_{2}t_{002}-t_{01}^{2}-t_{02}^{2}  \nonumber \\
&&+2t_{1122}-2t_{2}t_{112}+t_{12}^{2}-t_{0}^{2}t_{1}^{2})\ldots \,. 
\nonumber
\end{eqnarray}
The representation in the fundamental domain is $B_{2}$.
A detailed comparison between the semiclassical predictions 
and the exact quantum resonances for the 4-disk scattering problem has
been performed by \cite{Gaspard94}.

\section{Comparison with experimental resonances}

The experimental transmission function 
$|S_{21}(k)|^{2}=X^{2}(k)+Y^{2}(k)$ 
(Sec. II) can be expressed as a superposition of Lorentzian peaks,

\begin{equation}
|S_{21}(k)|^{2}=\sum_{i}\frac{c_{i}\gamma _{i}}{(k-s_{i})^{2}+\gamma _{i}^{2}
}\,,  \label{trace}
\end{equation}
where, as above, $s_{i}$, $\gamma _{i}$ denote respectively the position and
the half-width of the resonances in the $k$-domain. The parameters $c_{i}$
are coupling coefficients that depend on the location of the two probes and
reflect the coupling between the pick-up antenna with the $E_{z}$-pattern of
a given resonant mode. Semiclassical calculations using the appropriate
cycle expansion described in the previous section were performed for
different geometries, leading to the real and imaginary part $s_{i}$ and 
$\gamma _{i}$ of the resonances. In comparing with the observed traces,
the parameters $c_{i}$ were set manually to fit the data. For a given
scattering geometry, the trace $S_{21}$ used is the
average of several traces collected at different probe locations in order to
avoid missing resonances due to the accidental coincidence of either probe
with a node of the wavefunction. In general, good agreement 
is found for the resonant frequencies of
relatively sharp resonances (with typical quality factors in the range 
$Q=f/\Delta f\lesssim 50$). Broader resonances with larger imaginary parts
are instead not easy to distinguish, although all resonances are always
contributing to the transmission function. We examine now specific
configurations.

{\it 2-disk}: For the two-disk scattering, preliminary measurements were reported 
in \cite{Kudrolli95}. We carried out experiments in both the full and in the
half-space geometry, with $a=5$ cm and $R=40$ cm. According to the
discussion in Sec. IVa, the trace is expected to exhibit resonance peaks at
regularly spaced locations, $f_{n}=n$ GHz$,\,n=1,2,\ldots ,$ with a constant
width approximately equal to $\Delta f_{n}\simeq 0.29$ GHz (from Eq. 
(\ref{two})). A typical experimental trace is shown in Fig. 2, where we focussed
on the $A_2$-resonances between 0 and 20 GHz. The corresponding calculated trace 
is depicted as a dashed line. The agreement is found to be quite good
for both the resonances and their width. The regularity of such a spectrum will
manifest clearly in the corresponding autocorrelation function.

{\it 3-disk}: For the 3-disk geometry, we recall that a first demonstration of 
classical chaotic scattering via scattering of laser light was presented in 
\cite{Bercovich91}. A typical microwave trace for a 3-disk scatterer with 
$\sigma =4\sqrt{3}$ is presented in Fig. 3. Again, we focus on the fundamental 
domain representation of the scattering geometry, corresponding to resonances with 
$A_2$-symmetry. The semiclassical calculations, which are shown as a dashed line, 
are carried out by using the cycle expansion (\ref{ruelle2}) with 8 periodic orbits 
up to period 4. We verified that they accurately reproduce previous calculations 
on the same system \cite{Eckhardt93,Eckhardt95}. For this scattering geometry, 
comparison with the exact quantum mechanical calculations is also available 
\cite{Decanini98}, implying a stringent test for the validity of the semiclassical 
method. According to Fig. 3, the overall agreement is qualitatively good, especially 
for the locations of the sharper resonances. 

{\it 4-disk}: The traces of a 4-disk scatterer with $\sigma=4$ in the full space and 
the fundamental domain are shown in Fig. 4 top and bottom, respectively. 
Semiclassical calculations (dashed line) were performed by including a total of 14 
periodic orbits up to period 3 along the same procedure adopted in \cite{Gaspard94}. 
Resonances belonging to different symmetry characters can be identified in the 
full-space configuration and compared with the semiclassical predictions 
\cite{Gaspard94}. A similar comparison can be performed in the fundamental domain, 
only based on resonances in the $B_2$ subspace. 
As for the 3-disk case, the
semiclassical theory provides a qualitatively fair prediction of the 
resonance frequencies. We observe that, in general, the agreement for the 
lowest-energy widths is not as satisfactory, with discrepancies increasing with 
decreasing frequencies. This kind of discrepancy, which is also found in the 
3-disk geometry discussed above, is intrinsic to the semiclassical calculation 
because of the large correction of the stationary phase approximation \cite{Wirzba92}. 
For the 3-disk \cite{Gaspard89,Eckhardt95,Decanini98} and 4-disk \cite{Gaspard94} 
systems, where exact quantum-mechanical calculations are available, 
the very low-lying resonance widths of semiclassical resonances appear to 
be systematically smaller (up to a factor 3) compared to the corresponding 
quantum ones. 

A few general remarks are in order. Although the agreement between the 
experimental scattering resonances and the corresponding semiclassical 
predictions is generally within a few percents ($\approx 5\%$), some discrepancies 
are also evidenced from the data we analyzed. Such discrepancies may manifest in 
the form of both frequency shifts or width modifications of the predicted resonances 
as well as in the presence of additional peaks in the experimental trace. 
Various mechanisms and experimental limitations are expected to contribute as 
possible sources of errors, including: symmetry-breaking perturbations introduced 
by non-perfect geometries, effects associated with spurious reflections,  
non-idealities in the operations of the microwave absorbers 
({\it e.g.}, frequency-dependent response), or slight height variations over 
the cavity area. The combined action of such mechanisms makes 
open-geometry microwave experiments comparatively more demanding with respect to 
their closed-cavity counterpart, where some of the above error sources are 
practically irrelevant. While a deeper understanding of the unavoidable 
non-idealities faced by the experiments, along with the necessary technical 
improvements, are likely to be necessary for establishing a fully quantitative
detailed comparison, the level of agreement reached in our present investigation 
can be considered a very satisfactory match with the opportunity of 
retaining a relatively simple experimental methodology. 

\section{Spectral autocorrelation}

We now turn to analyze the data in terms of the so-called {\sl spectral
autocorrelation function}, which was calculated as

\begin{equation}
C(\kappa )=\left\langle \,|S_{21}(k-(\kappa /2))|^{2}\,|S_{21}(k+
(\kappa/2))|^{2}\,\right\rangle _{k\,\,}.
\label{average}
\end{equation}
Here, $\kappa $ is the wave-vector difference and $\left\langle
\,\,\,\right\rangle _{k}$ denotes an average over a band of wave-vectors
centered at some value $k=k_{0}$ and of width $\Delta k$, the latter being
large enough to include an appreciable number of resonances. The average
(\ref{average}) also includes a suitable window function which 
is chosen as \cite{Lai92} 
\begin{equation}
f(x)=\,\left\{ 
\begin{array}{ll}
1-|x|/\sqrt{6}\quad & |x|<\sqrt{6} \\ 
0 & |x|\ge \sqrt{6}
\end{array}
\,, \right. \quad x=\frac{k-k_{0}}{\Delta k}\,.
\end{equation}
Besides its intrinsic interest, an additional motivation for investigating
the properties of $C(\kappa )$ comes from the correspondence, pointed out in
Sec. IIIc, with experiments performed on mesoscopic transport. For ballistic
conductors, a formally similar magnetic-field correlation function received 
extensive theoretical and experimental attention as a potential
probe for quantum chaos \cite{Jalabert90,Lai92}. A similar autocorrelation
measure was also considered recently in the context of molecular
photodissociation spectra \cite{Alhassid98}. In our microwave experiments,
the wave-vector plays the role of the magnetic field and since 
$\,|S_{21}(k)|^{2}\propto T(k),$ the function $C(\kappa )$ can be regarded as
a measure of the wave-vector correlations of the two-probe conductance. The
dependence of the autocorrelation function on both the finite window 
$\Delta k$ and the center point $k_{0}$ has been checked in the calculations. We
consider the average of autocorrelations with a different $k_{0}$ to
compensate for slight dependences on the center point. Plots of typical
experimental autocorrelations for 2-, 3-, and 4-disk systems are shown in
Fig. \ref{fig5}, Fig. \ref{fig6} and Fig. \ref{fig7} respectively.

By inserting the explicit representation of $|S_{21}(k)|^{2}$ as a sum of
Lorentzians, Eq. (\ref{trace}), the autocorrelation is found as

\begin{equation}
C(\kappa )=\pi \sum_{i,j}\frac{c_{i}c_{j\,}(\gamma _{i}+\gamma _{j})}{
(\kappa -(s_{i}-s_{j}))^{2}+(\gamma _{i}+\gamma _{j})^{2}}\,.
\end{equation}
In a regime where there are no overlapping resonances, $|s_{i}-s_{j}|>>(
\gamma _{i}+\gamma _{j})$, and the small-$\kappa $ behavior of the
autocorrelation can be simplified as \cite{Eckhardt93}

\begin{equation}
C(\kappa )\approx \pi \sum_{i}\frac{2\,c_{i}^{2}\,\gamma _{i}}
{\kappa^{2}+4\gamma _{i}^{2}}\,.
\end{equation}
By exploiting a result from semiclassical Random Matrix Theory, the above sum
can be replaced by a single Lorentzian \cite{Blumel88,Lewenkopf91}: 
\begin{equation}
C(\kappa )=C(0)\,{\frac{1}{1+(\kappa /\gamma )^{2}}\,},  \label{lorentzian}
\end{equation}
where the parameter $\gamma =\gamma _{cl}$ is identified with the 
{\sl classical escape rate} from the chaotic scattering region, with the velocity
scaled to 1. Accordingly, one can interpret the width of the autocorrelation
function as an {\sl average width} (and thereby lifetime) of the resonances 
\cite{Ericsson60,Brink63}.

\subsection{Universal features: classical escape rate}

According to the above predictions, a {\sl universal }behavior of the the
autocorrelation function is expected for sufficiently {\sl small }
correlation scales, regardless of the details of the geometry and the way
the system is excited. Such universal behavior is captured by the single
classical parameter $\gamma .$ We recall its definition. Classically, if we
shoot particles towards the scatterer, the number $N(t)$ of particles
remaining in the scattering region after time $t$ decays exponentially as

\begin{equation}
N(t)=N(0)\exp (-\tilde{\gamma}_{cl}\,t)\,,  \label{clescape}
\end{equation}
where $\tilde{\gamma}_{cl}=\lambda (1-d)$ is the classical escape rate, 
$\lambda $ is the Lyapunov exponent of the manifold of infinitely trapped
orbits (strange repeller), and $d$ is the information dimension of the
unstable manifolds. The scaled escape rate, corresponding to unit velocity,
is defined as $\gamma _{cl}=\tilde{\gamma}_{cl}/v$, $v$ being the speed of
the particles. The classical escape rate can be calculated through the 
{\sl classical} Ruelle $\zeta $-function \cite{Gaspard92,Gaspard98},
\begin{equation}
\zeta_{\beta }(s)=\prod_{p}\,\left[ 1-(-1)^{L_{p}}\exp (sl_{p})/
\Lambda_{p}^{\beta }\right]^{-1},
\end{equation}
which is analytical in the half-plane 
${\rm Re} s<-P(\beta )$, and has poles in the other half-plane. In particular, 
$\zeta_{\beta }(s)$ has a simple pole at $s=P(\beta )$. Here, $P(\beta )$ is the
so-called Ruelle topological pressure, from which all the characteristic
quantities of classical dynamics can be derived in principle. The classical
escape rate is $\gamma _{cl}=-P(1).$

For the various scattering geometries investigated experimentally, we
calculated the appropriate autocorrelation function from the observed trace
and fitted the small-$\kappa $ portion of the resulting curve with the
Lorentzian behavior (\ref{lorentzian}), thereby extracting an
experimental escape rate $\gamma _{qm}.$ In general, good agreement is
observed with the classical prediction $\gamma _{cl}$, implying that in the
regime of universality {\sl the characteristic scale of wave-vector
correlations in the measured two-point quantum conductance is well
reproduced by knowledge of the chaotic classical scattering dynamics,}
through the classical escape rate $\gamma _{cl}$.

{\it 2-disk:} For the integrable two-disk system, the information dimension 
$d=0$, thus $\tilde{\gamma}_{cl}=\lambda $. For unit velocity and $R>2a$, we get 
\cite{Gaspard98}

\begin{equation}
\gamma _{cl}=\lambda =\frac{1}{R-2a}\,\ln \,\Lambda \,\,,
\end{equation}
$\Lambda \equiv \Lambda (\sigma )$ being the eigenvalue of the monodromy
matrix introduced in Eq. (\ref{Lambda}). The autocorrelation for the
experimental set up with $a=5$ cm, $R=40$ cm ($\sigma =8)$ is shown in Fig.
5. A value $\gamma _{qm}=0.083$ cm$^{-1}$ is found, which is in excellent
agreement with the classical result $\gamma _{cl}=0.088$ cm$^{-1}.$

{\it 3-disk:} For the three-disk scatterer, we have \cite{Gaspard98}
\begin{equation}
\gamma _{cl}\simeq \frac{1}{R}\ln \,(1.072\,\sigma \,)\,.  \label{escape3}
\end{equation}
A representative wave-vector autocorrelation function for this system
is displayed in Fig. 3, where the fundamental domain configuration
has been investigated with $a=5$ cm and $R=20\sqrt{3}$ cm. From equation 
(\ref{escape3}), $\gamma _{cl}\simeq 0.058$ cm$^{-1}$, to be compared with the
experimental value $\gamma _{qm}=0.064$ cm$^{-1}.$ The latter, leading to a
scaled value $\gamma _{qm}a=0.32,$ is in very good agreement with both
semiclassical and Monte Carlo estimates as given by \cite{Eckhardt95}. The
quantitative agreement between the small-$\kappa $ decay of correlations and
the Lorentzian curve demonstrates explicitly that behavior in such a region
is universal, with no dependence on the actual details of the geometry.

{\it 4-disk:} For the four-disk scatterer, we use the results of Refs.
\cite{Gaspard92,Gaspard98} for comparison with the experimental data.
Asymptotically, for large $R$, 
\begin{equation}
\gamma _{cl}\simeq \frac{1}{\sqrt{2}R}\ln \,(2\sqrt{2}\,\sigma )\,.
\label{escape4}
\end{equation}
The autocorrelation function for 4-disk data in the fundamental domain is
displayed in Fig. 7, corresponding to $a=5$ cm and $R=20\sqrt{2}$ cm.
The value $\gamma _{qm}=0.070$ cm$^{-1}$ compares pretty well with the 
estimate from Eq. (\ref{escape4}), $\gamma _{cl}=0.069$ cm$^{-1}$.

In Fig. 8, the experimental escape rates $\gamma _{qm}$ of the 4-disk
system are compared with
the classical prediction $\gamma _{cl}$ for several values of the ratio 
$\sigma$. Note that data are included for 17 configurations
of the different reduced ($1/8$, $1/4$, $1/2$ and full space)
representations of the 4-disk geometry shown in Fig. 4. The radius of the
disks used was $a=5$ cm for the $1/8$th space, and $a=2$ cm for the others.
A relevant quantity is the abscissa of absolute convergence $s_{c}$
for Eq. (\ref{ruelle2}), which can also be estimated from the Ruelle 
$\zeta$-function with the classical cycle weights $t_{p}$ replaced by the
corresponding semiclassical ones. $s_{c}$ serves as a crude lower bound of
the escape rate\cite{Eckhardt95}. The latter is also shown in Fig. 8. 
$s_{c}$ becomes negative for $R/a<4.5$.

By comparing the results found for the escape rate while passing from 2- up
to 4-disk scattering, progressively smaller values are obtained. In general,
it is interesting to examine the variation of the escape parameter with
increasing number $n$ of scatterers. For $n\rightarrow \infty$, one obtains
a so-called Lorentz scatterer \cite{Gaspard98}. We carried out experiments
with $n=20$, with corresponding (scaled) escape rate 
$\gamma _{qm}\simeq 0.05$, which is roughly an order of magnitude smaller 
than the three- or
four-disk values. In agreement with physical intuition, this indicates that
the system approaches a closed system when the number of disks becomes very
large. Accordingly, the escape rate from the chaotic region is found to be
quite small.

\subsection{Non-universal features}

For intermediate $\kappa $, the semiclassical prediction of 
Eq. (\ref{lorentzian}) fails because of the presence of the periodic orbits, 
which leads to {\sl non-universal} behavior. In the case of just one periodic
orbit, one may express the full two-point correlation function as 
\begin{equation}
C(\kappa )\propto \sum_{n=0}^{\infty }\frac{{2\gamma }}{(\kappa -n\Delta
s)^{2}+4\gamma ^{2}}\,,
\end{equation}
where $\Delta s$ denotes the spacing between resonances in the wave-vector
domain. For example, for the two-disk problem discussed above (Sec. IVa), 
$\Delta s=2\pi /(R-2a)$, provided the space is probed through the
antisymmetric $A_{2}$-representation only \cite{Kudrolli95}. Thus, the
autocorrelation oscillates with period $\Delta s$. Very good agreement is
found between experiment and theory for this integrable two-disk system
(Fig. 5), where the expected value $\Delta s$ $\simeq 0.21$ cm$^{-1}$ is
identical with the observed oscillation period.

For the three-disk scatterer in the fundamental domain, the average length
of the periodic orbit per period is roughly given by $(l_{0}+l_{1})/2$, the
mean separation of the resonances being therefore

\begin{equation}
\overline{\Delta s}_{FD}=\frac{4\pi }{2R-(2+\sqrt{3})a}\,.
\label{FD}
\end{equation}
The autocorrelation is expected to oscillate with a period roughly equal to 
$\overline{\Delta s}_{FD}$. The value $\overline{\Delta s}_{FD}=0.25\,$cm$^{-1}$ 
predicted from Eq. (\ref{FD}) is in very good agreement 
with the scale of the oscillations in Fig. 6.

Let us finally discuss the four-disk scatterer. In the full space, the
average length of the periodic orbits per period can be estimated as the
average length of the eight periodic orbits, $12$, $23$, $34$, $41$, $1234$, 
$1432$, $13$, $24$, where $1,2,3,4$ are the labels of the four disks \cite
{Cvitanovic93}. The mean separation between the resonances is then given
approximately by 
\begin{equation}
\overline{\Delta s}_{full}=\frac{2\pi }{(2+1/\sqrt{2})R-(3+\sqrt{2})a}\,.
\end{equation}
For the 4-disk system in the fundamental domain (one-eighth of the phase
space), the average length of the periodic orbit per period is 
$(l_{0}+l_{1}+l_{2})/3$, thus the mean separation is 
\begin{equation}
\overline{\Delta s}_{FD}=\frac{6\pi }{(3+\sqrt{2})R-2(2+\sqrt{2})a}\,.
\label{FD2}
\end{equation}
The autocorrelation will oscillate with an approximate period equal to 
$\overline{\Delta s}_{FD}$, which indicates the deviation from the 
semiclassical theory due to the presence of the periodic orbits. Thus, 
the large wave-vector (or short time) behavior is system specific. The value of 
$\overline{\Delta s}_{FD}=0.21\,$cm$^{-1}$ found from Eq. (\ref{FD2}) is in 
good agreement with the scale of oscillations in Fig. 7.

Non-universal contributions can play in general a crucial role in
determining the overall structure of the spectral autocorrelation, since
they can be of the same order of the universal result of Random Matrix
Theory. Semiclassical methods have recently provided an insightful tool
in modeling non-universal properties in addition to universal ones 
\cite{Agam98}. Beside the general remarks mentioned above, the systematic
identification of non-universal features, along with their interplay with
the universal scattering properties, deserves a separate experimental
investigation.

\section{\protect\smallskip Discussion and conclusions}

We presented an extensive experimental investigation of hard-disk
chaotic scattering in microwave open cavities. The experiments provide a
conceptually clean and direct realization of the $n$-disk open billiard
problem. By exploiting our (in principle unlimited) capability to vary the
geometry, chaotic scattering was studied in various configurations by
both changing the number of disks and the symmetry 
properties of the underlying phase space.

Two main conclusions can be drawn from the experiments. First, the general
validity and the predictive power of semiclassical methods have been tested
directly by comparing the observed spectra with the corresponding
semiclassical predictions. Qualitative agreement has been verified in all
the situations investigated, and quantitative comparison found for a 
wide class of relatively sharp resonances. 
In addition, the experiments point out the validity of the wave-vector 
autocorrelation function as a probe for phase-space structure and quantum 
chaos. Values of wave-mechanical escape rates have been extracted from the 
observed autocorrelations, and compared to their classical counterparts. 
In general, the agreement observed between
the measured escape parameters and the corresponding semiclassical
predictions tends to be quantitatively superior than the one reachable in
the detailed comparison of single resonance peaks. 
A similar conclusion has been reached for closed geometries where a much better
agreement between experimental and numerical results has been obtained for 
the statistical properties rather than for the comparison of the
individual resonances \cite{Alt99}.
This feature reflects the nature of the escape rate as an {\sl average} 
spectral property.

We remark that the present experiments, which probe wave-vector dependence, 
nicely complement measurements performed on semiconductor microstructures, 
where a similar role has been
stressed for the magnetic field correlation of the two-point conductance.
From the broader perspective of {\sl quantum-classical correspondence,} the
experiments shed light on the interplay between classical and quantum
features of the scattering dynamics, by showing that measurable properties
like quantum correlation lengths can be predicted from a knowledge of the
classical chaotic scattering behavior.

Our investigation clearly points out, among other issues, the need for a
deeper understanding of the non-universal properties of the spectral
statistics and their interplay with universal ones. Such an investigation is
likely to involve wave-vector autocorrelation functions as considered in the
present analysis, as well as different quantities useful to characterize and
probe the chaotic dynamics. In general, novel tools may be required in order
to pull out the whole amount of information encapsulated in the observed
spectra. Work is ongoing along these directions.

\section{Acknowledgments}

This work was supported by NSF-PHY-9722681.

\begin{figure}
\epsfig{width=.9 \linewidth,file=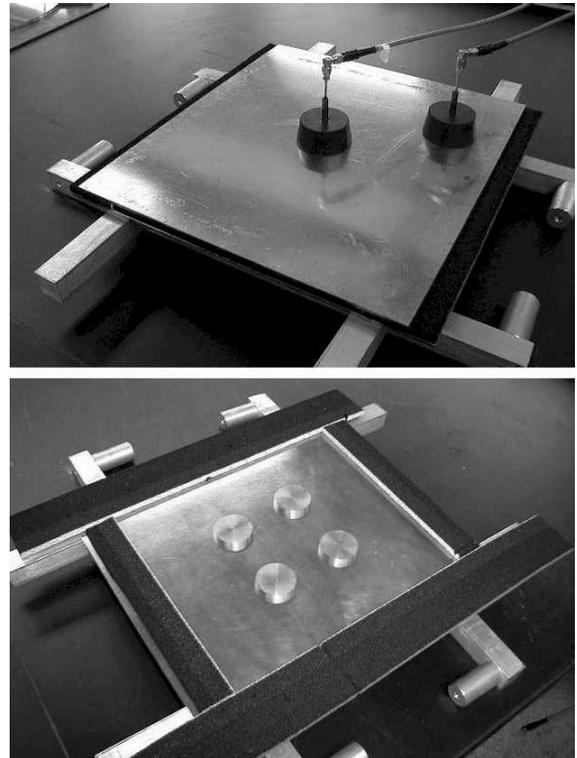}
\vspace{0.0truecm}
\vskip .4cm
\caption{ Photo of the experimental apparatus. (Top) Closed cavity. 
(Bottom) Open cavity. }
\label{fig1}
\end{figure}

\begin{figure}
\epsfig{width=.9 \linewidth,file=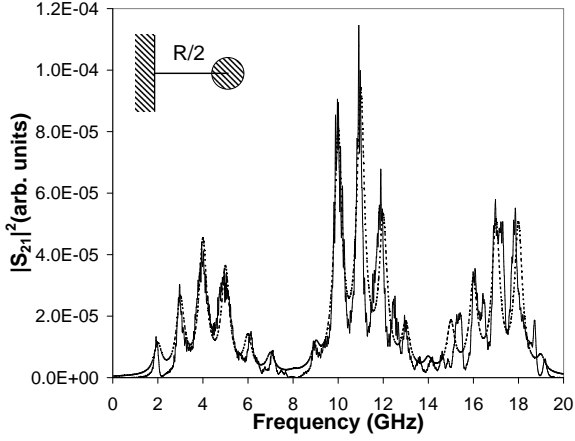}
\vspace{0.0truecm}
\caption{ Experimental (solid line) vs. semiclassical (dashed line) 
transmission function $|S_{21}|^2$ for a 2-disk system with 
$R=40$ cm and $a=5$ cm probed in the fundamental domain.
(Inset) Sketch of the corresponding experimental configuration.
The separation distance $R$ is indicated. }
\label{fig2}
\end{figure}

\begin{figure}
\epsfig{width=.9 \linewidth,file=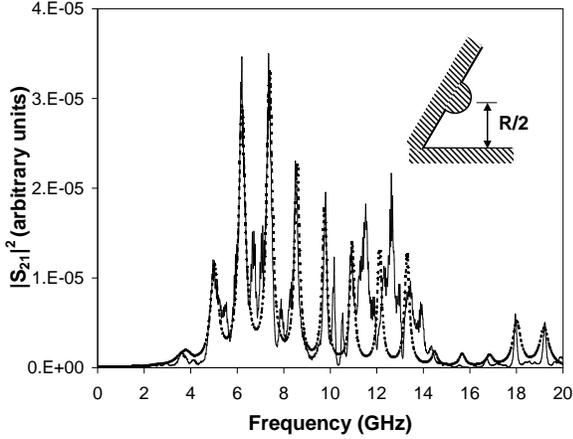}
\vspace{0.0truecm}
\caption{ Experimental (solid line) vs. semiclassical (dashed line) 
transmission function $|S_{21}|^2$ for a 3-disk system in the
fundamental domain with $R=20\sqrt{3}$ cm and $a=5$ cm. The corresponding 
experimental configuration is sketched in the inset. 
The separation distance $R$ is indicated.}
\label{fig3}
\end{figure}

\begin{figure}
\epsfig{width=.9 \linewidth,file=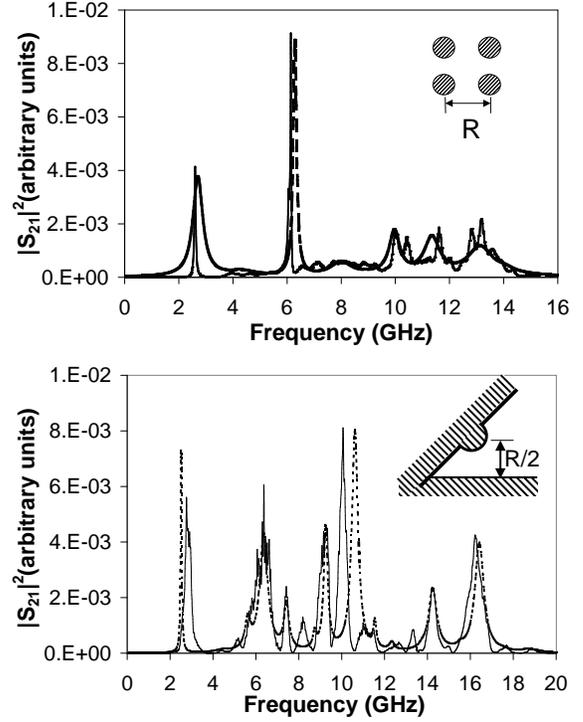}
\vspace{0.0truecm}
\caption{Experimental (solid line) vs. semiclassical (dashed line)
transmission function $|S_{21}|^2$ for a 4-disk system. 
(Top) Full space geometry with $R=8$ cm and $a=2$ cm.
(Bottom) 1/8th space (fundamental domain) with $R=20$ cm and $a=5$ cm.
The corresponding experimental configurations are sketched in the insets.
The separation distance $R$ is indicated.}
\label{fig4}
\end{figure}

\begin{figure}
\epsfig{width=.9 \linewidth,file=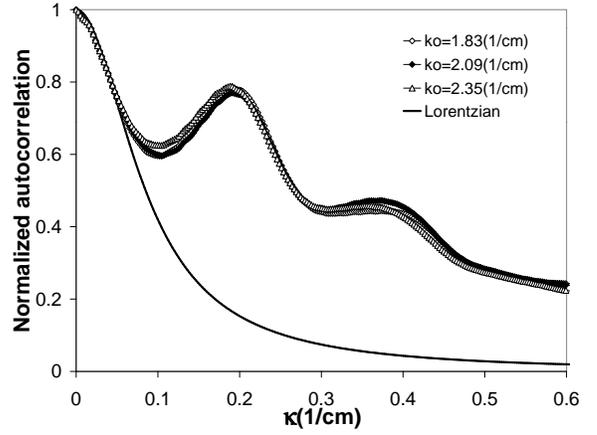}
\vspace{0.0truecm}
\caption{Wave-vector autocorrelation $C(\kappa)$ of the 2-disk system 
with $R$=40 cm and $a$=5 cm. Data are shown for the half-space configuration
of the 2-disk geometry, corresponding to the $A_2$ representation. 
The correlation is calculated with interval $\Delta k =3$ cm$^{-1}$. 
The different sets represent 
different values of the central wave-vector $k_0$. 
The bold line is a Lorentzian with $\gamma_{qm}$=0.083 cm$^{-1}$.}
\label{fig5}
\end{figure}

\begin{figure}
\epsfig{width=.9 \linewidth,file=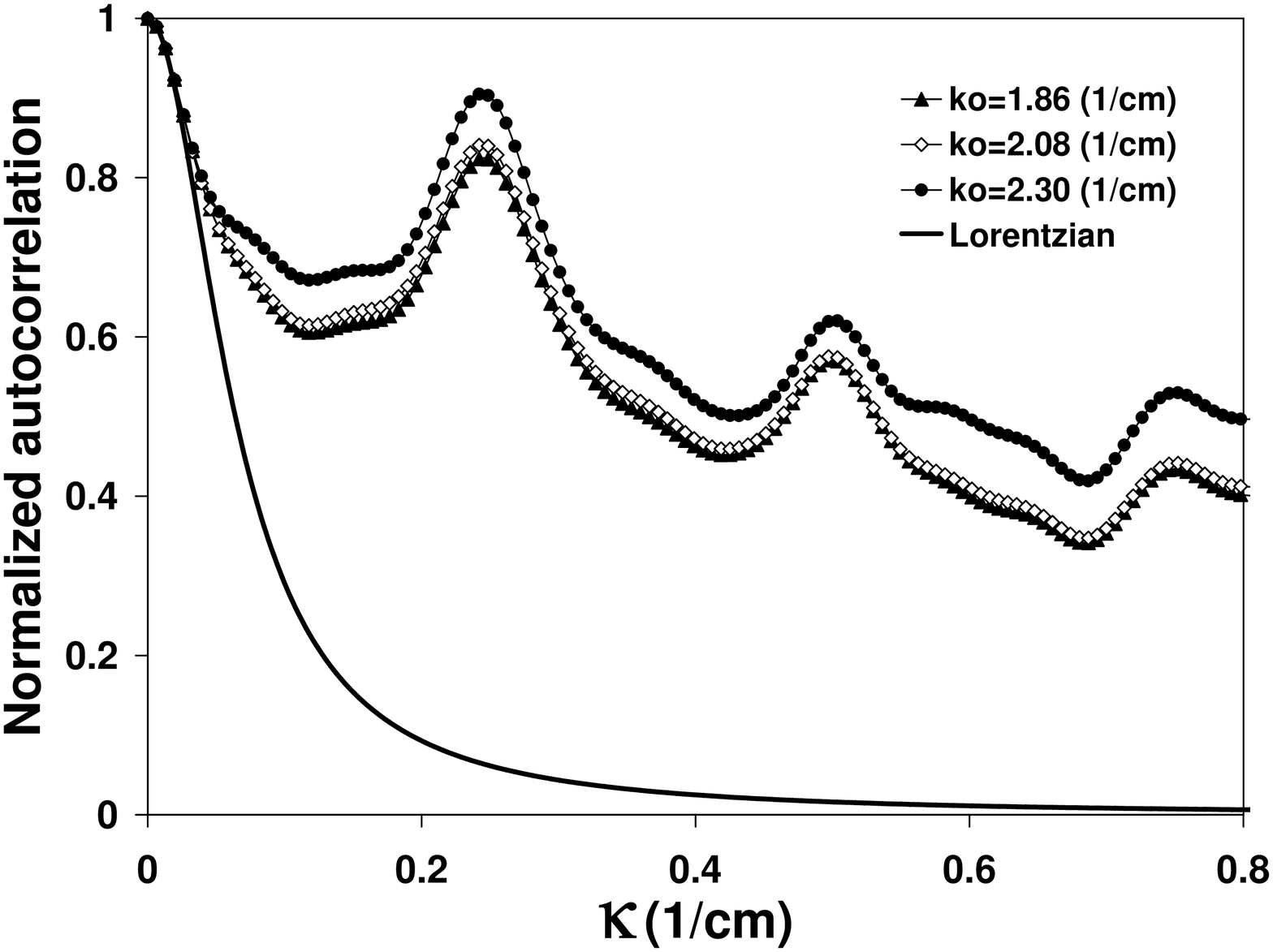}
\vspace{0.0truecm}
\caption{Wave-vector autocorrelation $C(\kappa )$ of the 3-disk system with 
$R=20 \sqrt{3}$ cm and $a=5$ cm. Data are taken in the fundamental domain, 
corresponding to the $A_2$ representation. The correlation is calculated with 
interval $\Delta k=2$ cm$\mbox{}^{-1}$. 
The different sets represent different values of the central wave-vector $k_0$. 
The bold line is a Lorentzian with $\gamma_{qm}=0.064$ cm$^{-1}$.}
\label{fig6}
\end{figure}

\begin{figure}
\epsfig{width=.9 \linewidth,file=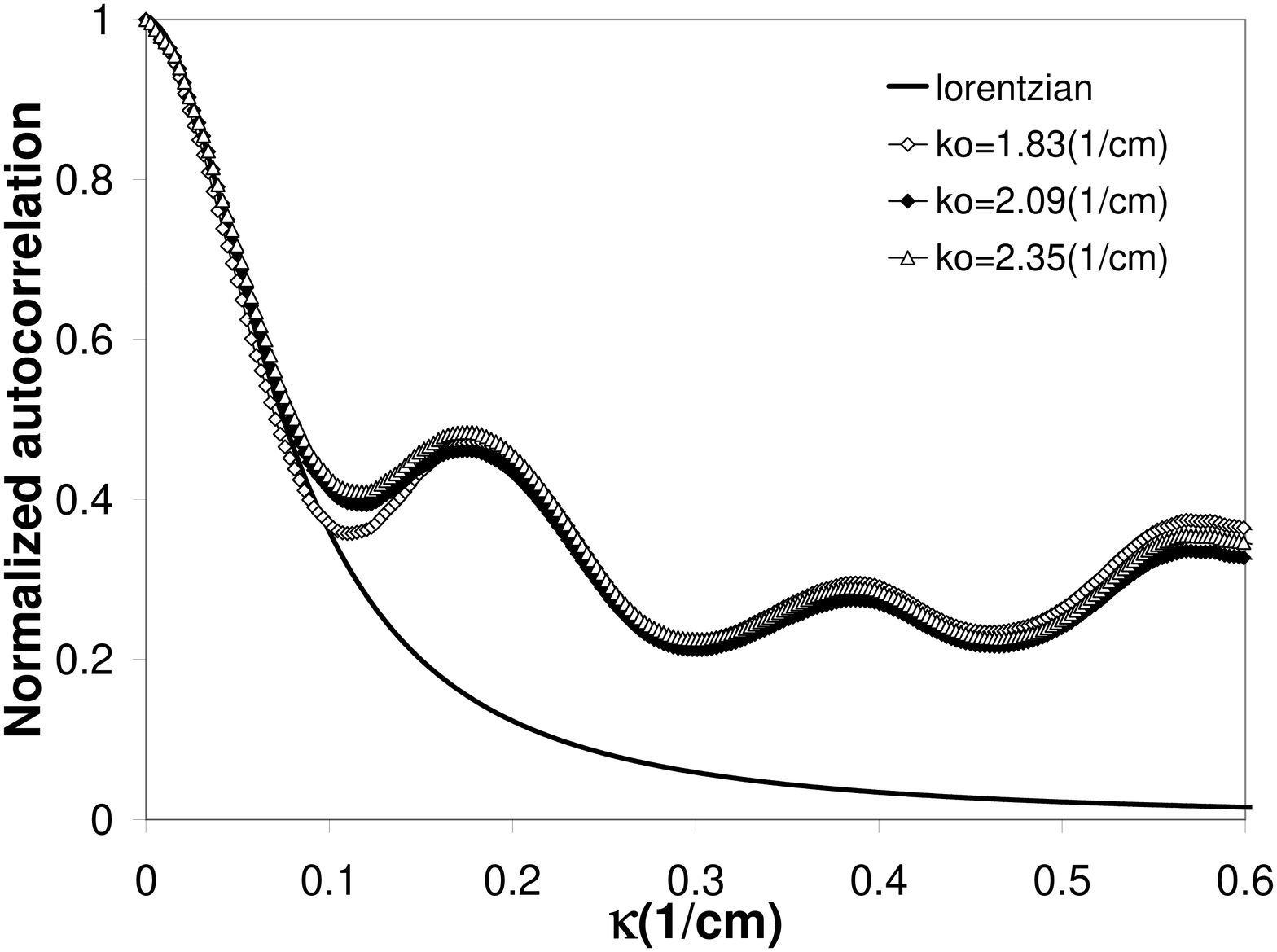}
\vspace{0.0truecm}
\caption{Wave-vector autocorrelation $C(\kappa )$ of the 4-disk system with 
$R=20\sqrt{2}$ cm and $a=5$ cm. Data are taken in the fundamental domain,
corresponding to the $B_2$ representation.
The correlation is calculated with interval $\Delta k=3$ cm$^{-1}$.
 The different 
sets represent different values of the central wave-vector $k_0$. 
The bold line is a Lorentzian with $\gamma_{qm}=0.070$ cm$^{-1}$. }
\label{fig7}
\end{figure}

\begin{figure}
\epsfig{width=.9 \linewidth,file=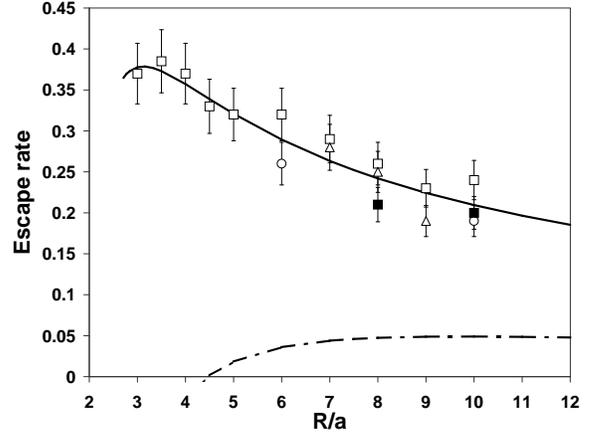}
\vspace{0.0truecm}
\caption{Experimental escape rate $\gamma _{qm}$ scaled to radius 
$a = 1$ vs. ratio $\sigma=R/a$. Data are shown for different 
reduced configurations of the 4-disk geometry: 
1/8th space (open squares), 1/2 space (open circles), 1/4th space (filled squares), 
full space (triangles).
The classical escape rate (solid line) is calculated from the 
first 3 periodic orbits in the fundamental domain. The abscissa 
of convergence $s_c$ of Eq. (17), which is shown as a dot-dashed line, 
represents a lower bound on the quantum escape rate.}
\label{fig8}
\end{figure}

\end{document}